\documentclass{article}

\usepackage{arxiv}

\usepackage[utf8]{inputenc} 
\usepackage[T1]{fontenc}    
\usepackage{hyperref}       
\usepackage{url}            
\usepackage{booktabs}       
\usepackage{amsfonts}       
\usepackage{nicefrac}       
\usepackage{microtype}      
\usepackage{lipsum}
\usepackage{graphicx}
\graphicspath{ {./images/} }
\usepackage{multirow}
\usepackage{amsmath}
\usepackage{adjustbox}
\usepackage{xcolor,xfrac}
\usepackage{graphicx}
\usepackage{tabularx} 
\usepackage{booktabs} 

\title{Modeling Changes in Individuals’ Cognitive Self-Esteem With and Without Access To Search Tools}

\author{
  Mahir Akgun \\
  College of Information Sciences and Technology\\
  Pennsylvania State University\\
  University Park, PA 16827 \\
  \texttt{makgun@psu.edu} \\
   \And
 Sacip Toker \\
  Information Systems Engineering\\
  Atılım University\\
  Ankara, Türkiye \\
  \texttt{sacip.toker@atilim.edu.tr} 
  \\
}

\begin{document}
\maketitle
\begin{abstract}
Search engines as cognitive partners reshape how individuals evaluate their cognitive abilities. This study examines how search tool access influences cognitive self-esteem (CSE)—users' self-perception of cognitive abilities—through the lens of transactive memory systems. Using a within-subject design with 164 participants, we found that CSE significantly inflates when users have access to search tools, driven by cognitive offloading. Participants with lower initial CSE exhibited greater shifts, highlighting individual differences. Search self-efficacy mediated the relationship between prior search experience and CSE, emphasizing the role of users’ past interactions. These findings reveal opportunities for search engine design: interfaces that promote awareness of cognitive offloading and foster self-reflection can support accurate metacognitive evaluations, reducing overreliance on external tools. This research contributes to HCI by demonstrating how interactive systems shape cognitive self-perception, offering actionable insights for designing human-centered tools that balance user confidence and cognitive independence.
\end{abstract}


\section{Introduction}
With improvements in information retrieval systems, access to extraordinarily large amounts of information has become easier, and so people engage in external information search more than ever. If we need to remember a statistical concept, learn how to solve a mathematical equation, or find out an answer to a general knowledge question, we can turn on our computers or smartphones, open up our web browser, and access what we are looking for immediately. As we are more experienced in finding information online, our reliance on web search tools increases, which leads us to form a memory partnership with search systems that maximizes both the amount of information available to us and the efficiency with which this information is stored and retrieved. The Internet has become a memory partner, and we tend to delegate the process of remembering to the Internet. The tendency to offload the responsibility for storing and retrieving information to technology has drastically affected our memory. As indicated by Sparrow et al. \cite{sparrow2011google}, if individuals believe they will have access to information on the internet, they are more likely to remember from where facts could be retrieved instead of remembering the facts themselves. It appears that as we become more reliant on the internet, we avoid our internal memory for information retrieval. 

\subsection{Internet and cognitive evaluations}

Besides the Internet’s impact on our ability to use internal memory to remember information, it has also changed how we evaluate the quality of our own memories. While access to the internet broadly facilitates cognitive offloading by providing vast repositories of information, the specific tools through which this occurs—namely, search engines like Google and Bing—play a more direct and nuanced role in shaping cognitive evaluations. Search tools are not merely gateways to the internet; they act as intermediaries that structure, filter, and present information in ways that actively influence how users perceive their cognitive capabilities. Some studies investigated how the use of web searching influences one's perceptions of their own cognitive abilities and showed that access to web search tools inflates one's belief in their cognitive abilities \cite{ward2013one, hamilton2018blurring}. Knowing that information being sought is readily available at any time on the Internet impacts metacognitive evaluations by increasing individuals’ confidence in their own memory and ability. As a type of cognitive evaluation, cognitive self-esteem (CSE) corresponds to one's perception of their own ability to think about, remember, and locate information \cite{ward2013one}. In the literature, for individuals who had access to the Internet or search tools, such as Google, higher CSE scores were reported, compared to those who had no access to the Internet or search tools. For example, Ward \cite{ward2013one} reported significantly higher CSE scores for individuals who used Google to answer a trivia question. Similarly, Hamilton and Yao \cite{hamilton2018blurring} revealed that people in the Google condition had significantly higher CSE scores than in both the no Google and control conditions. 

\subsection{Experience, self-esteem, and self-efficacy}
Self-esteem is a concept related to one’s belief about oneself. The relationship between experience and self-esteem has been studied in the relevant literature. Depending on the type of experience under investigation, some studies suggest a correlation, and others show a causal relationship between self-efficacy and experience. For example, Reitz et al. \cite{reitz2020self} showed that achievement-related experiences are correlated with self-esteem change among job beginners. Orth et al. \cite{orth2012life} indicated that self-esteem is a cause, rather than a consequence, of life outcomes, including relationship satisfaction, job satisfaction, occupational status, salary, positive and negative affect, depression, and physical health. On the other hand, Bachman and O’Malley \cite{bachman1977self} reported that occupational status-related experiences lead to increases in self-esteem. In a recent meta-analysis study, Krauss ad Orth \cite{krauss2022work} suggested a reciprocal pattern between self-esteem and work experiences. Previous research clearly shows that experiences and self-esteem are related, but the direction of the relationship depends on the type of experience under investigation. Given that the relationship between search experience and CSE, an aspect of self-esteem has been understudied, there is a need for research that helps us better understand such a relationship.

In addition to experience, self-efficacy is an important factor in improving self-esteem. Self-efficacy is one’s belief about their ability to complete a specific task  \cite{bandura1982self, cassidy2002developing, gardner1998self, yorra2014self}. Positive experience related to a task has a role in improving self-efficacy \cite{bailey2017measuring, cassidy2002developing, gardner1998self}. As an extension of what previous studies suggest, experience in information search can improve self-efficacy on future search tasks. Positive experience in finding information online improves one’s belief in their ability to access the information they need using online search tools. In several studies in different fields, including psychology, education, information systems, business, etc., the mediator role of self-efficacy has been repeatedly shown (e.g., \cite{abd2013role, appelbaum1996self,benight1999coping, brooke2017modelling, jashapara2011knowledge, nauta2004self, seo2008self, strobel2011yourself}). A mediated relationship between variables is plausible when those variables are shown as associated with each other and when there is one potential candidate as a mediator \cite{rm1986moderator, hayes2017introduction}. Given that search experience, self-esteem, and self-efficacy are associated, and that the mediator role of self-efficacy is known, self-efficacy can be hypothesized as a mediator of the relationship between search experience and CSE. Previous research demonstrated an inflated evaluation of internal knowledge when individuals had access to an external technological source; however, an individual’s perception related to his/her search experience has not been investigated in detail as a possible source of inflated evaluations. In the present study, we aim to explore the mediator role of self-efficacy in the relationship between search experience and CSE and whether this relationship is impacted by inflated cognitive evaluations that occur when an individual has access to external search tools.

\section{Theoretical Framework}
\subsection{The transactive memory system}
The theory of transactive memory was initially developed to explain transactive communication processes that facilitate access to information in human-human interactions. Partners of transactive memory (TM) systems are dependent on each other to know and remember specific content so that they do not have to remember content knowledge \cite{wegner1987transactive}. Rather, partners of such systems know where to locate the information, which helps them reduce the amount of information any given partner is supposed to remember \cite{wegner1995computer}. 

\subsection{Bridging Human-Computer Interaction, Transactive Memory Systems, and Cognitive Offloading}

 The concept of Transactive Memory Systems TMS describes how groups function as collective cognitive units, where individuals store and retrieve knowledge not only from their own memory but also from the expertise of others \cite{peltokorpi2008transactive}. Human-Computer Interaction (HCI) plays a central role in supporting this distributed memory by designing systems that make expertise visible and accessible. For instance, interactive platforms with expertise directories or knowledge visualization tools help users quickly identify who possesses the relevant information or skills to address specific tasks \cite{sarcevic2008transactive, ali2016social, gupta2018productivity}. These tools streamline collaboration and align with HCI’s emphasis on enhancing shared cognitive processes by creating interfaces that enhance human capability.

A critical factor influencing these systems is the ability to externalize cognitive tasks, a process known as cognitive offloading. Cognitive offloading occurs when users reduce their internal cognitive demands by relying on external systems, such as digital tools, to store or process information \cite{risko2016cognitive}. HCI supports this process by providing responsive and intuitive interfaces that reduce the perceived cost of accessing external memory systems. Research highlights that interface responsivity —how quickly and seamlessly information is retrieved— affects users’ decisions to offload. Highly responsive systems encourage externalizing memory by minimizing delays, while low responsivity deters offloading by increasing perceived effort \cite{grinschgl2020interface}. This dynamic illustrates the critical interplay between interface design and users’ cognitive evaluations of when and how to rely on external systems.

Interaction design also shapes offloading behavior by aligning interface mechanisms with users’ natural preferences. For example, touch-based interactions, perceived as more intuitive and effortless than mouse-based interactions, have been shown to increase cognitive offloading tendencies \cite{grinschgl2020interface}. Creating interfaces that match users’ established interaction patterns reduces cognitive barriers and fosters seamless integration of external memory systems into users’ workflows \cite{grinschgl2020interface}. 

The decision to offload cognitive tasks is not only shaped by interface design but also influenced by users’ metacognitive evaluations of cognitive costs and benefits. Wahn et al. \cite{wahn2023offloading} emphasized that metacognitive assessments play a central role in offloading, as users weigh their own cognitive capacity against the capabilities of external aids. For instance, under high cognitive load, individuals are more likely to offload attentionally demanding tasks to algorithms \cite{wahn2023offloading}. This willingness to offload reduces users’ cognitive burden and improves task performance. Such findings highlight the importance of designing interfaces that not only accommodate but actively support offloading behaviors in demanding contexts.

The structure and contextualization of information further play a vital role in supporting TMS and cognitive offloading. Context-sensitive user interfaces, which adapt dynamically to users’ tasks and environments, minimize cognitive effort by highlighting relevant information while filtering out distractions \cite{gauselmann2023relief}. By structuring information in a way that supports externalization, these interfaces facilitate offloading while preserving cognitive resources for the task at hand \cite{gauselmann2023relief}. The theoretical underpinnings of this relationship suggest that well-structured interfaces not only enable cognitive offloading but also shape users’ metacognitive evaluations of the costs and benefits of offloading in specific contexts.

Social networking platforms exemplify the integration of TMS and cognitive offloading principles in digital contexts. These platforms facilitate shared understandings of expertise and connections through persistent communication threads and visible networks \cite{oeldorf2020knows}. By drawing on these features, HCI enables collaborative tools that not only support group interactions but also allow users to offload cognitive tasks effectively, leveraging external systems as partners in distributed cognition.

Through these theoretical connections, it becomes evident that HCI bridges TMS and cognitive offloading by providing the design frameworks that make external memory partnerships feasible and effective.

\subsection{Cognitive offloading: the transactive memory partnership with the internet}

As a natural extension of human behavior, individuals may form TM partnerships with technologies. Previous research has proposed that people form a TM system with the Internet (e.g., \cite{fisher2015searching, sparrow2011google, ward2013one}) in which a human partner offloads memory to the Internet to reduce cognitive demand \cite{risko2016cognitive}. In such a TM partnership, the Internet stores all the knowledge and the human partner uses it for effortless external retrieval of the information \cite{fisher2015searching}. The human partner is never asked to retrieve information from internal memory, so the TM partnership with the Internet is considered one-sided. With the advent of search engines, people can offload more information than they would have stored internally to search tools that serve as their transactive memory. This transactive memory-based offloading enables us to shift from what information to remember to where information is located. Risko and Gilbert \cite{risko2016cognitive} argue transactive memory can impact metacognition. They also added that investigation of metacognitive aspects of the processes triggering cognitive offloading  (i.e., the use of physical action to reduce the cognitive demand of information processing of a task; \cite{risko2016cognitive}) and its outcomes may be beneficial for the theorization of cognitive offloading. They have proposed a metacognitive model of cognitive offloading where the selection of strategies based on either offloading (extended strategy) or reliance on memory (internal strategy) is dependent on the metacognitive evaluation of individuals’ mental capacities and external capacities with their body and world. As suggested by the model, the selection and successful engagement of an offloading strategy, resulting in positive consequences, favors the strategy, such as believing it is a more reliable source. As a result of the positive offloading experience, people do not tend to rely on their internal memory in future strategy selections. Such positive consequences (i.e., former cognitive evaluations; \cite{risko2016cognitive}) further influence subsequent metacognitive evaluations and cognitive processes. For instance, when one needs to prepare a dinner with some food alternatives for several guests, (s)he initially starts evaluating either using her/his unaided memory or searching on Google for available recipes. If (s)he has an opportunity to find different recipes and spend little time to prepare the main course food list via Google, (s)he does not need to memorize those recipes. When (s)he has a similar need again, (s)he will more likely rely on Google for the recipes instead of the internal memory. In this example, one’s first successful use of search engines as an offloading/extended strategy enabled them to rely on transactive memory, not needing to memorize the information for further recalls. The whole circular process of metacognitive evaluations, offloading information to transactive memory rather than internal memory, triggers the subsequent metacognitive evaluations and impacts the amount of information we want to load into our memory.

\subsection{Cognitive self-esteem: an aspect of self-esteem}

Self-esteem has different forms ranging from global to more specific concepts, such as academic ability \cite{coopersmith1982professional}, feeling about one’s own body \cite{franzoi1984body}, physical prowess \cite{fleming1990development}, professional self-esteem \cite{aricak1999group}, social identity \cite{luhtanen1992collective}, social skills \cite{fleming1990development}, etc. Ward \cite{ward2013one} asserted that even though there are different types of self-esteem, there was no construct directly depicting cognitive self-esteem, which was the most critical aspect of self-esteem when the blurring effect of the internet on individuals’ perception related to their memory is concerned. Cognitive self-esteem is composed of three components: ability to think, ability to remember, and transactive memory skills. These three components provide a better understanding of blurred evaluation of cognitive abilities by establishing links with the capabilities of the internet or a search engine. A search engine can give access to the thoughts and insights of others, record and deliver a vast amount of information, and, finally, index and seek information. With the presence of a search engine, individuals can perceive that they are better at thinking, remembering, and finding information. 

\section{Relevant Studies}

Studies in the literature highlight the emergence of search tools as a new memory partner that replaces a human partner as a companion in sharing the daily tasks of remembering. In an experiment conducted by Sparrow et al. \cite{sparrow2011google}, it was tested whether people remembered the information they expected to have later access better than the information they expected to be erased. In the experiment, participants read some memorable statements first and then typed them into the computer. Half of the participants believed that the computer would save what was typed, whereas half believed the statements they typed in would be erased. After the reading and typing task, participants were asked to write down the statements they recalled. The study showed that participants who thought the computer erased what they typed had a better recall performance than those who thought the computer saved what was typed. In another experiment, the researchers showed that participants remembered where to find the information better than the information itself. These findings suggest that new computing and communication technology is becoming a transactive memory that is evident when people seem better able to recall where an item is stored than the item itself \cite{sparrow2011google}.

The information we can access using the internet is much greater in scope than can be stored by any human partner; the internet provides us with a much easier and faster way to access up-to-date information. Sometimes, we can retrieve information from the internet more quickly than what we can retrieve from our own memory. Therefore, the boundary between the internal mind (i.e., what resides in our mind) and the external mind (i.e, what a partner knows) attenuates when a transactive memory system is formed with the internet. Ward \cite{ward2013one} investigated whether people incorporate the internet into a subjective sense of self. In the study, individuals were asked to answer trivia questions with or without using Google and then asked to rate themself on a CSE scale measuring how they assess the capability of their own memories. Ward's findings revealed that  CSE was significantly higher for people who used Google to complete a brief trivia quiz. Increases in CSE after using Google show that people perceive the internet as part of their own cognitive toolset \cite{ward2013one}. 

In another study, Hamilton and Yao \cite{hamilton2018blurring} investigated whether retrieving answers from Google would result in higher cognitive evaluations compared to retrieving answers from memory. In the study, participants were randomly assigned one of three conditions (Google, no Google, and control). In each condition, participants were asked to answer a ten-item quiz. Participants in the Google condition were allowed to use Google to find answers, whereas those in the no Google condition were asked to find answers on their own. Participants in the control condition were not instructed whether or not they should use the search engine. Immediately after completing the quiz, participants completed the CSE scale. Results revealed that people in the Google condition had significantly higher CSE scores than in both the no Google and control conditions. In addition, the study showed that using a familiar tool (e.g., search engines) or a familiar device (e.g., smartphone) to access information resulted in higher cognitive evaluations. Although the mechanism underlying the effect of the device or tool familiarity on evaluations of personal knowledge is not known well, Hamilton and Yao \cite{hamilton2018blurring} propose that accessing information through a familiar device may cause blurred boundaries between internal and external knowledge. 

On the other hand, Kahn and Martinez \cite{kahn2020text} explored the Google effects on memory and CSE in the context of Snapchat and text messaging. “Google effect” occurs when someone is less likely to remember information in the event that they believe they have future access to it. Given that Snapchat messages are more ephemeral than text messages, the researchers hypothesized that individuals who received Snapchat messages as a source of information would have a better memory and higher CSE than those who had information presented as text messages. Unlike previous studies, Kahn and Martinez \cite{kahn2020text} reported no memory and CSE differences between the two channels. 

\section{Present Study}

In line with the relevant literature, when designing the study, we presumed that (1) people provide lower cognitive evaluations when they do not have access to search engines, compared to the situation in which they can access such tools, (2) in a sequence of evaluations, people’s former evaluations can impact latter evaluations, (3) self-efficacy has a mediator role in the relationship between search experience and CSE. To operationalize our thoughts related to CSE, search experience, and search self-efficacy, we formulated three hypotheses for the present study: 
\begin{itemize}
    \item  H1: The CSE of participants is lower when they do not have access to search engines but higher when they have access to search engines, 
    \item  H2: The change in CSE that occurs over time varies depending on participants' former cognitive evaluations and
    \item  H3: The search self-efficacy can mediate the association between search experience and CSE. 
\end{itemize}
   
\section{METHOD}   
\subsection{Participants}
 We conducted this study in two undergraduate courses in the same semester at a large northeastern university in the United States. The research was approved by the institutional review board and consent was obtained from students in two courses. Participation was voluntary and was not compensated. Following the completion of obtaining consent from students, we collected data in the first three weeks of the semester. For students to be included in the study, they were expected to (a) consent to participate in the study, and (b) complete all tasks assigned to them. 202 students enrolled in the courses; 164 students (22.6 \% female) consented to participate in the study and completed tasks in all experimental conditions. Therefore, the sample size for the study is 164.

\subsection{Experimental Design}
Ward \cite{ward2013one} indicated that when the Internet is perceived as a transactive memory, individuals’ belief about their performance is blurred between intrinsic ability (e.g., knowledge) and situational influences (e.g., access to a search engine). They suggest that the impact of the situational factor causing a blurred effect on self-evaluation of internal memory can be revealed by giving individuals similar tasks in two consecutive conditions: condition 1, in which the situational cause is absent, and condition 2, in which the situational cause is present. Ward \cite{ward2013one} also suggests the use of general knowledge questions in the consequent conditions to examine the impact of situational factors on cognitive evaluations. In line with these suggestions, we designed no access and access conditions, assuming that participants could tend to answer a question from their internal memory in the no access (NA) condition and that they tend to use a search engine to answer the question in the access (A) condition. In NA conditions, the participants were not allowed to use any device to access search engines. In contrast, in A conditions, they were permitted to use search tools to search for the general knowledge questions. 

Participants were asked to respond to two consecutive general knowledge questions. Each question was presented initially when students had no access to search tools (e.g., Google or Bing) and then re-presented when they had access to search tools. The purpose of asking the same general knowledge question twice (i.e., first in the NA condition, then in the A condition) was to minimize potential impacts of asking different questions on CSE in NA and A conditions and to better examine the change in CSE that occurred when the participants gained access to search tools. Given that the study aims to examine a potential change in participants’ CSE scores, we needed at least three temporarily separated observations to examine such a change. Since we could not use the same question after participants found the correct answer in the A condition, there was a need to use a similar task to create a third observation. Therefore, we used two different general knowledge questions with the same difficulty. 

\subsection{Selection of general knowledge questions}
Tauber et al. \cite{tauber2013general} produced the general knowledge questions to assess the illusions related to false memory recalls and investigate individuals’ confidence judgments, an index about the certainty of the accuracy of their own knowledge. Moreover, these questions were developed for assessing confidence and peer metacognitive judgments \cite{tauber2013general}. Specifically, Tauber et al. \cite{tauber2013general} calculated the probability of recalls for each question, which assists researchers to choose questions with a probability value very close to zero to ensure that it is very unlikely for individuals to answer them from memory. For these reasons, we used two of these questions in this study, and for similar reasons, these questions were used in other studies (e.g., \cite{ferguson2015answers, hamilton2016judging}). According to Hamilton et al. \cite{hamilton2016judging}, CSE is a form of global judgment about self rather than the judgment of domain-specific knowledge, so general knowledge questions appear to better serve the purpose of assessing CSE.

We selected general knowledge questions with the recall probability of .000 (i.e., very close to zero). Asking difficult questions helped us understand how the participants’ self-evaluations could change across all experimental conditions when answering questions, they did not know. If the participants had already known the answer to these questions, they would not have needed to use a search engine in a condition.

The questions selected for the study are presented below:

\begin{itemize}
    \item Q1: What is the highest mountain in South America?
    \item Q2: What is the name of the instrument used to measure wind speed?
\end{itemize}

\subsection{Experimental Conditions}
Each participant was exposed to access and no access conditions in the order presented below:
\begin{itemize}
    \item Answer a general knowledge question when having 'no access to search tools'  (NA)
    \item Answer a general knowledge question when having 'access to search tools' (A)
    \item Answer a second general knowledge question when having 'no access to search tools' (NA)
    \item Answer a second general knowledge question when having 'access to search tools' (A)

\end{itemize}

We applied a counter-balancing approach \cite{gaito1961repeated} to minimize the ordering effect of the within-subject design - one of the significant internal validity threats. We created two combinations of the general knowledge questions presented to the participants by changing the order of questions (see Table 1) during the experiment. The order of “no-access” and “access” was constant for both of the questions. Group 1 (G1) responded to Q1 first and then Q2. Group 2 (G2) saw Q2 first and then Q1. For data analysis, the participants from both groups were included, so the sample size remained at 164. To ensure the observed effects were due to the intervention, we implemented strict controls. Participants were randomly assigned to counterbalanced conditions to minimize order effects and completed tasks in a controlled environment where search tool access was regulated. Time intervals (2, 3, and 2 days between sessions) allowed participants to reset their cognitive state between tasks while reducing confounding influences from daily search activities. 

\begin{table}[h!]
\centering
\caption{Counter-balancing of the general knowledge questions} 
\begin{tabular}{c c c c c}
\hline
\textbf{Group} & \textbf{C1} & \textbf{C2} & \textbf{C3} & \textbf{C4} \\ \hline
\textbf{G1}    & GQ1-NA      & GQ1-A       & GQ2-NA      & GQ2-A       \\
\textbf{G2}    & GQ2-NA      & GQ2-A       & GQ1-NA      & GQ1-A       \\
\hline
\end{tabular}
\end{table}

\subsection{The Instruments}
Three measures were used: cognitive self-esteem (CSE), search self-efficacy, and search experience.

\subsubsection{Cognitive Self-Esteem (CSE)}
The measure for Cognitive Self-Esteem (CSE) was developed and validated by Ward \cite{ward2013one} to assess self-perceptions of cognitive abilities, including memory, reasoning, and problem-solving. The CSE scale is a 14-item Likert scale with response options ranging from Strongly Disagree to Strongly Agree. Ward’s study established the measure’s discriminant validity, demonstrating low correlations with related constructs such as general self-esteem and self-efficacy (r < .50) and strong factor loadings for unique dimensions of cognitive self-esteem. Explanatory factor analysis revealed three factors underlying the CSE construct: (1) confidence in the ability to think, (2) confidence in the ability to remember, and (3) transactive memory skills. Factor loadings for these items ranged from .60 to .90, supporting the construct validity of the scale. Cronbach’s $\alpha$ as reliability evidence for the overall scale and factors ranged from .78 to .94, indicating satisfactory internal consistency.

The test-retest reliability of the measure was reported as .82 over a two-week period in Ward’s validation study, supporting the stability of CSE for short-term interventions. To further ensure reliability in the present study, Pearson Product Moment correlation analysis was conducted between the CSE scores for the access and no-access conditions of the two general knowledge questions. The correlation between the CSE measures for Q1 (access vs. no-access) was r = .846, and for Q2 (access vs. no-access) it was r = .792, indicating strong stability across conditions and satisfactory test-retest reliability within the present study’s context.

In this study, Ward’s measure was adopted to investigate the situational influences of search tool access on participants’ cognitive self-esteem. By incorporating a well-validated instrument, we ensured that the CSE construct was distinct from other related concepts such as general self-esteem and self-efficacy, as evidenced by prior discriminant validity analysis (r < .50) and supported the reliability of the measure through factor loadings, Cronbach’s $\alpha$, and test-retest correlations.

\subsubsection{Search Self-Efficacy}

The search self-efficacy scale was developed to measure individual personal judgment related to searching for action for performing \cite{brennan2016factor}. The items of the scale were developed from several studies related to information search tasks. The data from 327 participants were collected to run the explanatory factor analysis. Principal axis factoring and oblique rotation (Promax) were utilized. The items loading ranged from .626 to .818, and the analysis generated four factors: (1) overall task success, (2) effective use of time, (3) query development skills, and (4) advanced search skills. The questionnaire uses 10-point scale, where 1 = Totally unconfident, 5-6 = Reasonably confident; 10 = Very confident.  
\subsubsection{Search Experience}
The search experience measure was a single-item measure that asks to what extent participants can express their expertise as a user from beginner or novice, intermediate, advanced, and expert \cite{bates1990should, smith2015domain, wildemuth2004effects}. The item used in the search experience measure is on an interval scale.  

\subsection{Procedure}
Data were collected in four sessions in two information technology courses that met twice a week in person. Before the first session, participants were randomly distributed to one of the groups presented in Table 1. In the first session (C1), before the lecture, the first general knowledge question (Q1 or Q2) was presented to students in each group; then, students were asked to answer the question from memory.  Then, students completed the questionnaires explained in the previous section. All questionnaires were distributed online via Qualtrics. In the second session (C2), before the lecture, the second general knowledge question (Q1 or Q2) was presented to students in each group; then, students were asked if they knew the answer to the question. Students who did not know the answer were asked to complete the questionnaires. Students who knew the answer were not allowed to complete the task in C2, so they were excluded from the study. The same process was repeated in C3 and C4. Figure 1 shows the procedure followed in the study. 

This study measured transient effects on cognitive self-esteem (CSE) over a short timeframe to capture immediate, task-related influences. The session spacing (2, 3, and 2 days) was strategically chosen based on prior research \cite{tauber2018does}. These short intervals helped minimize external influences while maintaining the impact of the intervention. The overall focused intervention period of one to two weeks was designed to provide a balance, allowing sufficient time for participants to process their experiences while ensuring the continued influence of the intervention. Research indicates that self-esteem and related cognitive constructs can change in short periods, supporting the notion that changes in CSE can occur within the time frame of this study. For example, Zuffian et al. \cite{zuffiano2023relation} observed daily changes in self-esteem and emotional self-efficacy over 10 days. Similarly, Dan et al. \cite{dat2022effectiveness} demonstrated that self-esteem could be enhanced through targeted interventions, even when implemented over short durations. Lishinski and Yadav \cite{lishinski2021self} further showed that self-evaluation interventions could improve both self-efficacy and performance in educational settings. Collectively, these studies establish that self-esteem, self-efficacy, and other cognitive self-assessments are sensitive to short-term experiences and interventions, aligning with the one-to-two week structure employed in the present study.

While a longer-term design could provide deeper insights into the lasting effects of search tool access, the timeframe adopted in this study is consistent with prior work examining context-dependent cognitive constructs \cite{ward2013one, hamilton2018blurring}. By limiting the study to a one-two-week intervention, this research aimed to ensure that the changes in CSE were attributable to the intervention itself rather than being confounded by participants’ routine search tool use or other daily experiences. This careful balance enabled a focused examination of how situational experiences dynamically shaped participants' CSE across the multiple sessions.

In this study, former cognitive evaluation is operationalized in line with the Cognitive Off-Loading Model \cite{risko2016cognitive}. CSE in the first condition is the former evaluation of CSE in the second condition; CSE in the second condition is the former evaluation of CSE in the third condition; CSE in the third condition is the former evaluation of CSE in the fourth condition. 

\begin{figure}[h!]
    \centering
    \includegraphics[width=\textwidth]{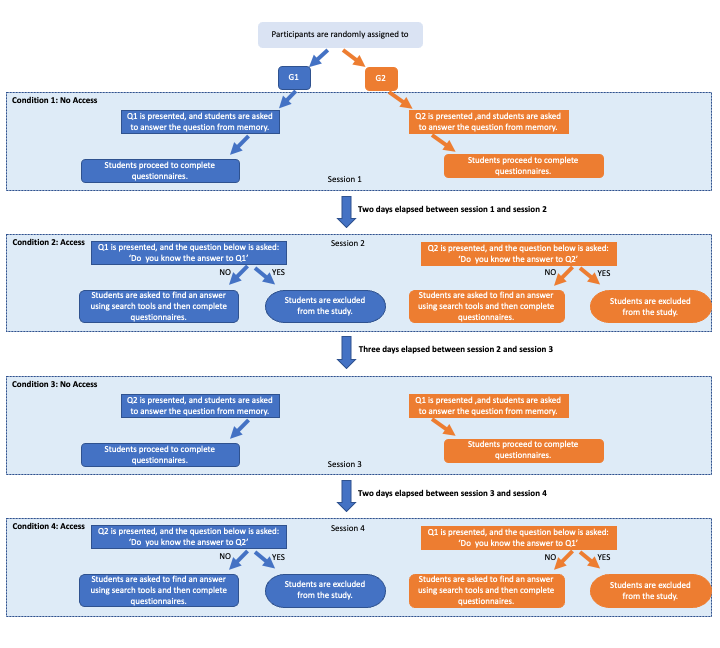} 
    \caption{The procedures of the study} 
    \label{fig:figure1} 
\end{figure}

\subsection{Data Analysis}
The main motivation of this study was to explain the change that occurred in the participants' CSE across four subsequent conditions. General Linear Models (e.g., paired sample t-test, ANOVA, ANCOVA, and repeated-measures ANOVA) do not provide sufficient information for us to explain how change occurs over time in relation to how it can interact with other factors
\cite{acskar2009ortuk, bereiter1963some, duncan2002latent, xitao2005power, lohman1999minding, muthen1998longitudinal}. The Latent Growth Model (LGM) is often preferred over linear models, especially when (1) change is the main focus of analysis, because of its capability to reveal individual differences between people for the variables investigated in a repeated-measure conditions, (2) there is an apriori hypothesis that bases on previous research \cite{park2005introduction, jung2008introduction, serva2011using}, (3) both individual and group levels of changes are investigated \cite{park2005introduction,jung2008introduction, serva2011using}, and (4) the main focus of a study is to reveal the associations among individuals and classify those individuals into distinct groups based on their response patterns \cite{muthen2000integrating}. In this study, LGM was preferred over general linear models because (1) we focused on the individual differences of CSE in repeated conditions, (2) we had theory-driven hypotheses, and (3) we measured one variable at four different time points to observe a change in CSE. Building on the strengths of Latent Growth Modeling (LGM) for analyzing individual differences and group-level patterns, this study extends the framework to Latent Growth Change (LGC). By explicitly modeling both initial levels and rates of change, LGC provides deeper insights into the dynamics of individual growth across repeated conditions.

In LGC, the analysis of mean and covariance structures in structural equation modeling is used to explain individual growth \cite{mcardle1987latent, meredith1990latent}. LGC includes an intercept (i.e., an individual's score on outcome variable at the onset of the investigation)  and slope (i.e., the growth or change rate at which outcome measure changes over time) parameters that jointly define within-person patterns of change. That latent structure helps researchers to understand the growth curve in terms of observed and latent variables by handling both modeling and estimation of measurement error \cite{byrne2013structural}.  

The factor loadings of intercept are equal to one in LGC models; since it is assumed that intercept remains constant across all occasions of measurement times. For slope, the loadings are identified based on the ‘change’ trend, which can be linear, quadrant or cubic \cite{heck2013multilevel}. Aşkar and Yurdugül \cite{acskar2009ortuk} recommend the use of a simple line graph that shows the change to observe the type of change. Since we had more than 100 participants, we created average scores of CSE for each condition to observe the type of change.

\begin{figure}[h!]
    \centering
   \includegraphics[width=\textwidth]{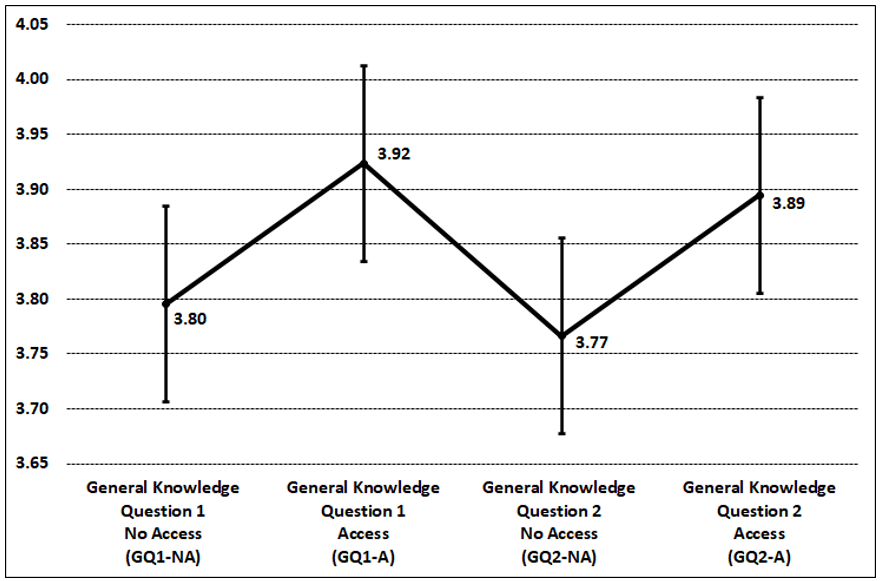} 
    \caption{Line graph of all participants’ average CSE scores for each condition} 
\end{figure}

Figure 2 shows how the average CSE scores change across four experimental conditions in a descriptive manner. The results indicate a cubic relationship between CSE scores and the conditions. Participants' CSE scores were at the lowest level in GQ1-NA, and they started increasing in GQ1-A. A decrease was observed in GQ2-NA that preceded an increase in CSE scores in GQ2-A. For this type of slope, the use of cubic constrained parameters for Slope loading ( i.e., 1-0-1-0) is suggested \cite{heck2013multilevel}. Therefore, in this study, we used slope coding of 1, 0, 1, 0 instead of the traditional 0, 1, 2, 3 in the latent growth model to capture the cubic trend observed in participants’ CSE scores. This non-linear coding reflects the alternating increases and decreases across the four conditions, unlike linear coding, which assumes consistent additive change. The 1, 0, 1, 0 coding assigns weights that better model these oscillatory changes, aligning the analysis with the observed data pattern. This approach follows methodological guidance advocating flexible slope coding for non-linear trends \cite{byrne2013structural, asparouhov2006multilevel}. In line with this suggestion, we used 1, 0, 1, 0 slope loadings values on observed variables. Moreover, we started our analysis with an unconditional model to understand within-person change by analyzing the relationship between intercept and slope. 

Later, we incorporated search experience into our unconditional model to understand if search experience helps us explain the random variability at the onset of the investigation and the rate of change. We added search self-efficacy to the unconditional model as a second predictor representing detailed skills. Including both search experience and self-efficacy in the unconditional model extended the conditional model.

We conducted a mediation analysis to explore the potential impact of search self-efficacy on the relationship between search experience and CSE. In some cases, a mediation variable stands for full mediation, meaning that the causation variable cannot directly impact the outcome variable when the mediator was present. To examine mediation, Barron and Kenny \cite{rm1986moderator} suggest checking the following four steps: 

\begin{enumerate}
    \item The causation variable must be significantly associated with the outcome variable
    \item The mediator variable must be significantly associated with the outcomes variable
    \item The causation variable must be significantly associated with the mediator variables
    \item Under the provision of the mediator variable, the causation variable must not be significantly associated with the outcome variable. 

\end{enumerate}

When all these steps are approved, it can be inferred as evidence for mediation. In the current study, search experience was a causation variable; search self-efficacy was a mediator variable; CSE was an outcome variable. We examined the potential impact of search self-efficacy on the relationship between search experience and CSE by using the four-step approach. Both LGC models and mediation analyses were examined in IBM SPSS AMOS 24.0 using bootstrapping. 

\section{Results}
\subsection{Order Effect}

We checked the difference between these groups and CSE scores and did not find a significant difference. This indicated that no ordering effect was present for the order of the questions, F(3, 483) = .285, p = .789. Since the order of questions did not have a critical impact, we reported the results based on the order of conditions.

\subsection{Unconditional Model}
The model presented in Figure 3 was tested to reveal whether or not (1) the CSE of participants in ‘no access to search engine’ conditions is lower than that of participants in ‘access to search engine’ conditions (H1), and (2) the change in CSE that occurs over time varies depending on participants' former cognitive evaluations (H2). First, we looked for a good model fit and then interpreted means, variances, and estimates.   

\begin{figure}[h!]
    \centering
   \includegraphics[width=\textwidth]{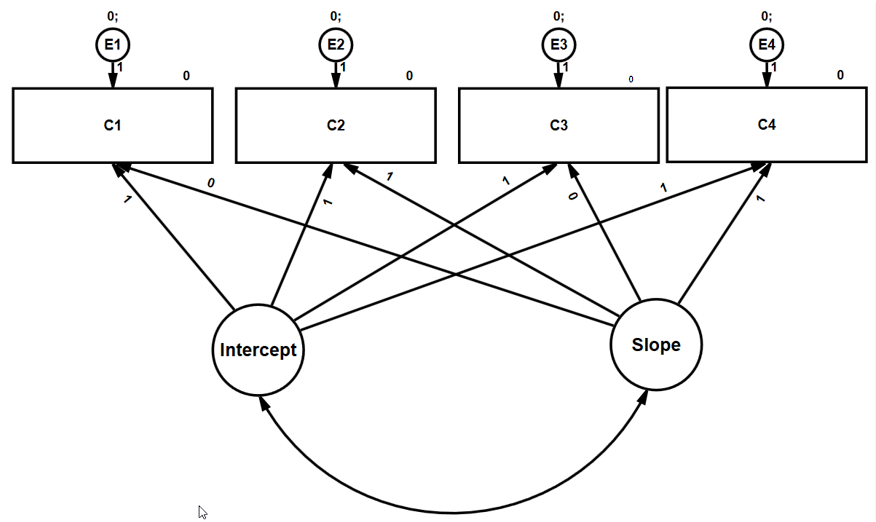} 
    \caption{The Latent Growth Model of CSE for H1 and H2} 
\end{figure}

\begin{table}[h!]
\centering
\caption{Evaluation of Unconditional LGC Model Fit Indices}
\begin{tabularx}{\linewidth}{X X X X X}
\hline
\textbf{Fit Index} & \textbf{Initial Model Value} & \textbf{Final Model Value (After modifications)} & \textbf{Criteria for Perfect Fit} & \textbf{Resource} \\ \hline
$\chi^2$ (df) & 59.914 (df = 3) p < .01 & 1.023 (df = 3) p = .796 & Low $\chi^2$ value and p > .05 & \cite{hooper2008structural}) \\
$\chi^2$/df & 19.971 (df = 5) & .341 (df = 3) & $\chi^2$/df < 3 & \cite{wheaton1977assessing,tabakhnick2007using}\\
RMSEA & .341 & .001 & RMSEA < .05 & \cite{hu1999cutoff, steiger2007understanding}\\
SRMR & .0460 & .0014 & SRMR $\leq$ .05 & \cite{byrne2013structural, diamantopoulos2000introducing} \\
GFI & .863 & .999 & .95 $\leq$ GFI $\leq$ 1 & \cite{tabakhnick2007using, miles2007time} \\
AGFI & .543 & .999 & .85 $\leq$ AGFI $\leq$ 1 & \cite{tabakhnick2007using}\\
CFI & .889 & .999 & .97 $\leq$ CFI $\leq$ 1 & \cite{hu1999cutoff} \\
IFI & .890 & .999 & .95 $\leq$ IFI$\leq$ 1 & \cite{miles2007time}\\
NNFI & .885 & .999 & .97 $\leq$ NNFI $\leq$ 1 & \cite{hu1999cutoff, fan1999effects}\\

\hline
\end{tabularx}
\end{table}

When the first model was run, the model fit values did not produce a good fit (see Table 2). Byrne \cite{byrne2013structural} suggested that when model fit does not produce good fit results, that may happen due to misspecification of the model, so modification indices can be taken into consideration in LGC models. For that reason, modification indices were applied for a better model specification. As can be seen in Figure 4, we included covariances both between E1 and E2 and between E3 and E4. As a result, a good fit was obtained with the standardized estimated model illustrated in Figure 4 (see Table 2 for model fit indices). This result indicates a change in CSE that occurred across four conditions in cubic form. The average CSE of participants decreased in GQ1-A, showed an upward trend in GQ2-NA, and finally went down again in GQ2-A. Stated differently, the total score of CSE when participants had access to search tools was higher than the total score when they had no access. This result provided supporting evidence for our first research hypothesis; regardless of what general knowledge questions were answered by participants, in ‘no access’ conditions, participants’ CSE was lower, compared to their CSE in ‘access’ conditions.

\begin{figure}[h!]
    \centering
   \includegraphics[width=\textwidth]{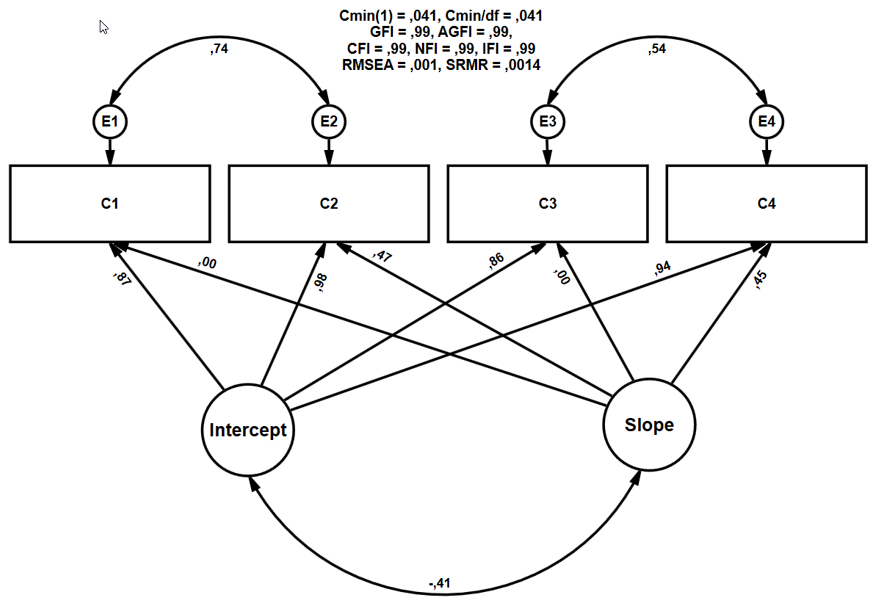} 
    \caption{Estimated LGM of CSE} 
\end{figure}

The modified model provides estimated means and variances for both the slope and intercept. The mean score estimated for the intercept is 3.78 (S.E. = .044, t = 80.587, p < .001), which indicates that the participants had a medium level of confidence in their own ability to remember answers to the general knowledge questions presented at the beginning of this investigation. The variance estimated for the intercept is .131 (S.E. = .024, t = 5.528, p < .001), which points out that there were individual differences in CSE at the onset of the investigation. The mean and variance of the slope are .272 (S.E. = .036, t = 7.626, p < .001) and .060 (S.E. = .012, t = 5.206, p < .001), respectively, indicating that the change was not the same for all participants across all conditions. The covariance between intercept and slope was -.053 (S.E. = .015, t = -3.541, p < .001), which illuminates that those with low CSE demonstrated more changes than those with high CSE, as illustrated in Figure 5. Based on these results, we divided the participants into two groups (i.e., the low CSE group and the high CSE group), statistically applying a two-step cluster analysis. The analysis revealed two distinctive groups with .99 Silhouette measure of cohesion and separation. The ratio of the largest and smallest groups was 1,83. In the low CSE group, there were 58 participants (35.4\%), while there were 106 participants (64.6\%) in the high CSE group. A comparison of the demographic characteristics of the high and low CSE groups revealed minimal differences. The high CSE group (n = 106) consisted of 23.6\% female participants, with an average age of 20.3 years (SD = 1.8). The low CSE group (n = 58) included 20.7\% female participants, with an average age of 20.1 years (SD = 2.0). Academic backgrounds were evenly distributed across both groups, with no significant differences observed in participants’ familiarity with search tools or prior experience with similar tasks (p > .05). Similarly, the demographic composition of participants in the two experimental conditions (access vs. no access) was consistent across groups. Gender, age, and search experience did not significantly differ between the conditions (p > .05), indicating that the random assignment effectively minimized potential demographic bias. Participants with high CSE scores in C1 condition where they had no access to search tools demonstrated a slight increase in CSE scores after they gained access to search tools (C2). In contrast, participants with low CSE scores in C1 condition showed a notable increase in CSE scores after they gained access. The same pattern was observed between C3 and C4, which indicates that the pattern was independent of the question participants were asked to answer. To ensure this CSE change is not associated with students’ prior knowledge of the general knowledge questions, we analyzed the responses for ‘no access’ conditions in Table 3. Table 3 demonstrates that the number of correct responses was not different for two questions across the groups. The unconditional model supports the second hypothesis (H2) of the present study.

\begin{table}[h!]
\centering
\caption{Crosstabs of the frequencies of Incorrect and Correct responses to both general knowledge questions}
\begin{tabularx}{\linewidth}{X X X X}
\hline
\textbf{Questions} & \textbf{Responses} & \textbf{CSE Level Low} & \textbf{CSE Level High} \\ \hline
General Knowledge Q1\textsuperscript{a} & Incorrect & 94 & 60 \\
                                       & Correct   & 5  & 5  \\
General Knowledge Q2\textsuperscript{b} & Incorrect & 96 & 57 \\
                                       & Correct  & 3  & 8  \\
\hline
\end{tabularx}
\footnotesize{Note. Cochran-Armitage test for trends in 2x2 crosstabs for General Knowledge Question 1 was not significant, CA = .477, p = .491; while General Knowledge Question 2 was significant, CA = 5.396, p = .020. \\
\textsuperscript{a} Due to counterbalancing, the frequency of responses at C1 for Group 1 and C3 for Group 2 (See Table 1 for more details) are presented. \\
\textsuperscript{b} Due to counterbalancing, the frequency of responses at C1 for Group 2 and C3 for Group 1 (See Table 1 for more details) are presented.}
\end{table}

Figure 5 also reveals the growth and time relationships in descriptive level with average scores. In the high CSE group, there were not many changes observed between conditions; moreover, a decrease was observed. Accessing a search engine did not make a significant fluctuation in CSE scores. On the other hand, in the low CSE group, they showed a constant increase when search engines were accessible. The inflated evaluation of CSE was getting deeper, and it would have ended up with a replacement of what was really known internally with what the search engine knew. Overall, when individuals had high CSE, accessing search engines and possible blurred effects of transactive memory tools did not appear impactful. However, when individuals had low CSE, they were more open to getting influenced by search engines related to the evaluation of their knowledge.

\begin{figure}[h!]
    \centering
   \includegraphics[width=\textwidth]{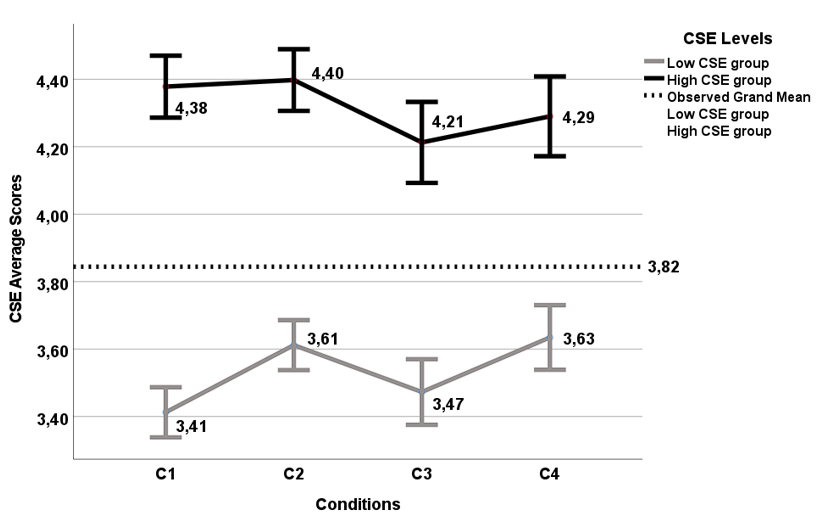} 
    \caption{High and Low CSE Groups Average Score Change Comparison  } 
\end{figure}
\subsection{Conditional Model with Search Experience and Search Self-Efficacy}
To test H3, once the optimal baseline growth model was established as an unconditional model, the model was expanded as a conditional model to include search experience and search self-efficacy as predictors of each condition and the change (see Figure 6). We examined whether these two variables explain the change in CSE that varied across all conditions depending on participants' CSE at the onset of the investigation. According to Brennan et al., \cite{brennan2016factor}, the search self-efficacy is expected to surpass the impact of the search experience, and this led us to check a possible mediation impact \cite{hayes2017introduction}, which required further analysis.

\begin{figure}[h!]
    \centering
   \includegraphics[width=\textwidth]{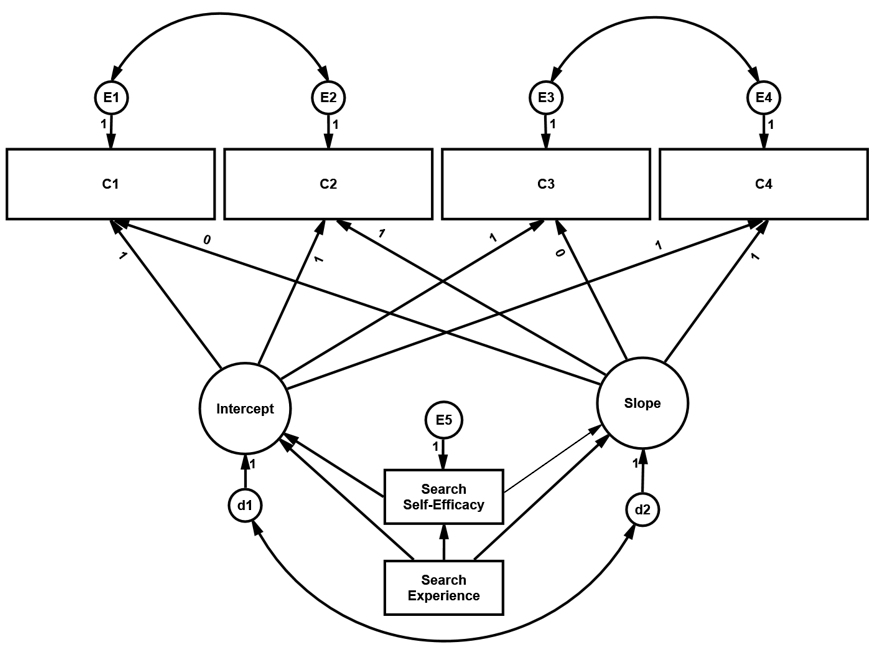} 
    \caption{The Conditional LGM Model of CSE with Search Experience and Self-efficacy for H3 } 
\end{figure}
Following the steps suggested by Barron and Kenny \cite{rm1986moderator}, the associations estimated by bootstrapping presented below were checked for a possible mediation:

\begin{enumerate}
    \item Between search experience and intercept of unconditional model: There was a significant impact of the search experience on the Intercept of LGM (B = .192, $\beta$ = .249, t = 3.027, p < .01),
    \item Between the search self-efficacy and intercept of unconditional model: There was a significant impact of the search self-efficacy on the Intercept of LGM (B = .134, $\beta$ = .414, t = 4.685, p < .001),
    \item Between search experience and search self-efficacy: There was a significant impact of the search experience on the search self-efficacy (B = 1.154, $\beta$ = -.482, t = 7.017, p < .001),
    \item Between search experience and intercept of unconditional model under the presence of search self-efficacy: Search experience had no significant impact on the intercept under the presence of search self-efficacy (B = .038, $\beta$ = .049, t = .557, p = .578). 
\end{enumerate}

For the slope, the same steps were checked; however, the first step did not yield a significant association between search experience and the slope, (B = -.008, $\beta$ = -.023, t = -.240, p = .881); the second step did not show a significant association between search self-efficacy and the slope as well (B = -.007, $\beta$ = -.047, t = -.430, p = .667). 

These results together indicate that there was a full mediation of search self-efficacy on the relationship between search experience and the intercept of the unconditional model with 77.5\% explained variance ratio. There was no significant impact of either search experience or search self-efficacy on the slope, so there was no need to check the mediation of search self-efficacy on the relationship between search experience and the slope of the unconditional model. The estimations of the conditional model are illustrated in Figure 7, and the model fit values for the model are presented in Table 4.

\begin{table}[h!]
\centering
\caption{Evaluation of Conditional LGC Model Fit Indices}
\begin{tabularx}{\linewidth}{X X X X}
\hline
\textbf{Fit Index} & \textbf{Model Value} & \textbf{Criteria for Perfect Fit} & \textbf{Resource} \\ \hline
$\chi^2$ (df) & 6.377 (df = 5) p = .271 & Low $\chi^2$ value and p > .05 & \cite{hooper2008structural}\\
$\chi^2$/df & 1.275 (df = 5) & $\chi^2$/df < 3 & \cite{wheaton1977assessing,tabakhnick2007using} \\
RMSEA & .041 & RMSEA < .05 & \cite{hu1999cutoff, steiger2007understanding}\\
SRMR & .0201 & SRMR $\leq$ .05 & \cite{byrne2013structural,diamantopoulos2000introducing}\\
GFI & .987 & .95 $\leq$ GFI $\leq$ 1 & \cite{tabakhnick2007using, miles2007time}\\
AGFI & .947 & .85 $\leq$ AGFI $\leq$ 1 & \cite{tabakhnick2007using} \\
CFI & .998 & .97 $\leq$ CFI $\leq$ 1 & \cite{hu1999cutoff}\\
IFI & .998 & .95 $\leq$ IFI $\leq$ 1 & \cite{miles2007time}\\
NNFI & .989 & .97 $\leq$ NNFI $\leq$ 1 & \cite{hu1999cutoff, fan1999effects}\\

\hline
\end{tabularx}
\end{table}

\begin{figure}[h!]
    \centering
   \includegraphics[width=\textwidth]{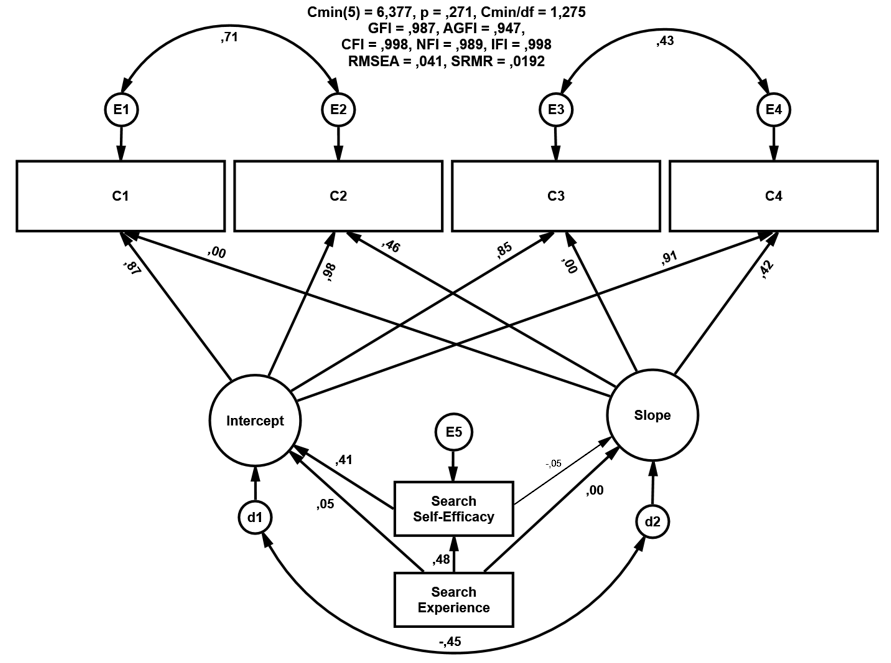} 
    \caption{Estimated conditional LGC model with Search Experience and Self-Efficacy Mediation } 
\end{figure}

The indirect unstandardized, standardized effects, and their confidence intervals of both search experience and search self-efficacy on CSE in each condition for overall, low and high groups were calculated by using the bootstrapping technique and presented in Table 5. For the overall scores, search experience and search self-efficacy did not have any significant direct or indirect impact on the slope of LGM. These results suggest that these two variables along with their mediation had an isolated impact on CSE in each condition, whereas they did not influence the change in CSE across four conditions. Our results showed that while search experience and search self-efficacy significantly influenced the intercept (C1–C4 levels), they had no significant impact on the slope of the LCM. This indicates that these variables shaped baseline and condition-specific levels of CSE but did not affect the overall rate of change across conditions. This suggests that search experience and self-efficacy primarily stabilize or elevate CSE within each condition without altering its trajectory over time.

For the individual differences, we found in the unconditional model, we analyzed low and high CSE groups’ estimates of only search self-efficacy due to its mediation role in the search experience. The low group yielded significant indirect estimates in all conditions; in contrast, the high group did not demonstrate any significant estimates. Furthermore, the estimates of the low group were higher than the high group. For all conditions, regardless of the low or high group, the estimates of access conditions were higher than ‘no access’ conditions. These results show the impact of search self-efficacy on CSE.

\begin{table}[h!]
\centering
\caption{Indirect Unstandardized and Standardized Estimates of Searching Experience and Self-Efficacy on CSEs}
\begin{tabularx}{\textwidth}{X X X X X}
\toprule
& \multicolumn{4}{c}{\textbf{Unstandardized B, Standardized ($\beta$) [Lower Bounds, Upper Bounds]}} \\ 
\cmidrule(lr){2-5}
 & \textbf{Search Experience (Total)} & \multicolumn{3}{c}{\textbf{Search Self-Efficacy}} \\ 
\cmidrule(lr){3-5}
 & & \textbf{Total} & \textbf{Low} & \textbf{High} \\ \midrule
\textbf{Slope} & -.008, -.023 \newline [-.038, .023] & - & - & - \\ 
\textbf{Intercept} & .155*, .199 \newline [.074, .245] & - & - & - \\ 
\textbf{C1} & .193*, .217 \newline [.053, .301] & .134*, .361 \newline [.068, .200] & .069*, .254 \newline [.019, .115] & .050, .247 \newline [-.001, .119] \\ 
\textbf{C2} & .185*, .234 \newline [.077, .279] & .127*, .382 \newline [.059, .184] & .078**, .292 \newline [.032, .126] & .052, .249 \newline [-.002, .125] \\ 
\textbf{C3} & .193*, -.217 \newline [.053, .301] & .134*, .353 \newline [.068, .200] & .069*, .222 \newline [.019, .115] & .050, .153 \newline [-.001, .119] \\ 
\textbf{C4} & .185*, .234 \newline [.077, .279] & .127*, .356 \newline [.059, .184] & .078**, .241 \newline [.032, .126] & .052, .179 \newline [-.001, .125] \\ 
\bottomrule
\end{tabularx}
\footnotesize{Note. Dashes indicate zero value. * p < .05, ** p < .01.}
\end{table}

\section{DISCUSSION}
Developments in the area of search and information access technologies have opened up ways for people to access the extraordinarily large amount of information and to engage in external information search more than ever. There is a growing tendency among people to rely on technology as a source of information instead of their own memory. Previous research suggests that people in a human-computer partnership, in which humans can delegate the responsibility for storing and retrieving information to technology, have inflated evaluations about the quality of their own memories. 

The present study provides supporting evidence for the position that using any systems, enabling cognitive offloading, impacts individuals’ cognitive evaluations (H1). An important finding of this study is the inflation of cognitive self-esteem (CSE) in the presence of search engines, driven by transactive memory systems. While search engines provide valuable cognitive support, they can lead to an overestimation of one’s cognitive abilities, fostering a reliance on external tools. This inflation of CSE may have unintended consequences, such as overconfidence in decision-making or reduced intrinsic motivation to acquire knowledge independently. For example, students may perform better on tasks requiring immediate recall when search tools are present but fail to develop deeper learning strategies for long-term retention and critical thinking. This finding aligns with previous research (e.g., \cite{ward2013one, wegner2013google}) that highlights how the distinction between what individuals know and what the search system knows can blur within a human-computer transactive memory system. The inflation of CSE observed in our study may further amplify this blurring effect, as users increasingly conflate their cognitive abilities with the capabilities of the search tools they rely on.  While this blurring of knowledge boundaries is evident, it may not affect all users equally. Our study reveals that individual differences in baseline CSE influence how much users' self-perceptions shift when they gain access to search tools (H2). The changing pattern in CSE varies depending on participants’ CSE at the onset of the investigation (H2). Participants who had relatively higher CSE demonstrated very little change in their CSE once they gained access to search tools (i.e., from GQ1-NA to GQ1-A, and GQ2-NA to GQ2-A), whereas those with relatively lower CSE demonstrated a noticeable change in their CSE after gaining access. From the onset to the end of the investigation, participants with higher CSE demonstrated a slight decrease in their CSE; those with lower CSE exhibited a noticeable increase in their CSE. This observation highlights the dynamic nature of cognitive self-esteem, which appears to fluctuate in response to contextual factors. Temporary boosts in CSE may enhance confidence and engagement, but they could also contribute to variability in task persistence and problem-solving strategies. The stability of these effects, however, remains an open question. The test-retest reliability of .78 over the study period suggests moderate short-term stability, but it is unclear whether repeated exposure to search tools produces cumulative or lasting effects on CSE. While these fluctuations are evident, they do not affect all users equally. Our findings are limited in their ability to explain why being part of a human-technology transactive memory system has a subtle effect on subsequent cognitive evaluations of individuals with high CSE, while subsequent cognitive evaluations of individuals with low CSE are impacted to a greater extent. Risko and Gilbert’s metacognitive model of cognitive off-loading \cite{risko2016cognitive} allows us to consider a possible explanation. According to Risko and Gilbert, our knowledge regarding our previous success with either our internal storage (i.e., our confidence level in our own mind) or external storage (i.e., our beliefs about the reliability of a digital memory partner) plays a role in deciding whether we store information internally or offload memory onto our digital partner. Based on the model Risko and Gilbert \cite{risko2016cognitive} propose, we speculate that the individual difference we found in the present study can be related to differences in decisions on whether or not to offload the responsibility of retrieving information to technology. Individuals who had relatively lower CSE throughout the study may have seen search tools as reliable partners that could assume the role of retrieving information. The beliefs they have regarding the reliability of the search tools may blur the boundary between their mind and the search tools as a source of retrieving knowledge, which, in turn, may lead to inflated CSEs after they gain access to search tools, such as Google. On the other hand, the other group of participants may have had relatively higher CSEs in the study because of their high confidence in their own internal memory. These individuals may tend to rely more on their internal memories than search tools as a source of knowledge, and therefore they may be more accurate in evaluating what they know and what their partner knows in the partnership they formed with search systems. This can be the reason why we did not observe much change in their CSE even after they had access to search tools. 

While the transactive memory framework provides a robust theoretical explanation for the observed effects, it is important to consider alternative perspectives that may further clarify shifts in cognitive self-esteem (CSE) during search tasks. One possibility is that shifts in CSE were driven by task difficulty or performance expectations. Tasks completed with search tools may have appeared easier, leading participants to infer higher cognitive ability based on perceived success. In contrast, tasks without tool access might have increased cognitive load, resulting in temporary reductions in CSE. Another potential explanation involves self-serving attribution biases. Participants may have attributed task success in the tool-access condition to their own cognitive abilities, while attributing failure in the no-tool condition to external constraints. This attributional pattern could artificially inflate CSE in the presence of tools, independent of transactive memory dynamics.

While these alternative accounts offer valuable perspectives, the transactive memory framework remains a robust explanation for the observed changes, as it directly addresses the interplay between external cognitive aids and self-perception. Future research could experimentally isolate these factors to better understand their relative contributions.

We can extend the scope of how individuals' decisions influence their CSEs to the role of search self-efficacy in such decisions. Our results indicate that the impact of search self-efficacy on CSE varies depending on the levels of CSEs that individuals demonstrated throughout the study. Search self-efficacy had a significant impact on CSEs for individuals who demonstrated lower levels of CSEs in each study condition. This impact increased when those individuals gained access to search tools in GQ1-A and GQ2-A. Search self-efficacy is based on people’s beliefs about their overall task success, effective use of time, query development skills, and advanced search skills \cite{brennan2016factor}. Since these beliefs are shaped by individuals’ prior successful search experiences, individuals’ beliefs about their abilities to accomplish tasks using search tools may lead to inflated CSEs once they gain access to search tools. Individuals’ prior search experience may lead to increased confidence in search systems, which may, in turn, influence their decision to offload memory onto the search systems as their digital partner.  On the other hand, CSEs for individuals who consistently demonstrated higher levels of CSEs were not influenced significantly by search self-efficacy; this insignificant effect remained nearly the same across all study conditions. We think that individuals with higher levels of CSEs at the onset of this investigation may see their own memory as a primary source of retrieving information because of their knowledge regarding their previous success with internal memory; they may use search tools if they believe their own memory does not have the information they seek. Their tendency to consult their own memory in the first place may limit the role of search self-efficacy in their cognitive evaluations.	

Several studies proposed concepts that are relevant to search experiences, such as device familiarity \cite{hamilton2018blurring} and search fluency \cite{stone2021search}. Investigating search self-efficacy and its role in cognitive evaluations in relation to these concepts may give us a better understanding of why search self-efficacy influences individuals’ CSEs differently over time depending on the level of their CSEs at the onset. For example, Hamilton and Yao \cite{hamilton2018blurring} demonstrate that the familiarity a user has with a device plays an important role in influencing cognitive evaluations. In our study, some individuals used their personal devices, and others used classroom computers. Taken together, cognitive evaluations of individuals who used personally owned devices may have been inflated by the device familiarity. Therefore, we suggest future research to control the device familiarity when investigating the impact of search self-efficacy on CSE.

Brennan et al. \cite{brennan2016factor} argue that search self-efficacy is an overarching measurement for search experience and provides more enriched data. Our study provides supporting evidence for this argument; we found that search self-efficacy highly mediated the association of search experience on CSE. An implication of this finding is to replace widely used single-item search experience measures with the search self-efficacy scale to measure search experience in future studies. 

The present study suggests that it is crucial to determine confounding factors that impact cognitive evaluations and control them when examining individuals' cognitive evaluations in a transactive memory structure they form with an external digital memory partner. Our findings indicate that search self-efficacy has an impact on CSE measured before each search task presented in the study, so it should be controlled or considered in future studies to better understand the underlying mechanisms of the impact of external digital memory partners on individuals’ cognitive evaluations. 

\subsection{Implications}
Implications from these results contribute to our understanding of how cognitive evaluations are influenced when individuals form a transactive memory system with an external digital partner.  The presence of individual differences at the beginning of a search task played a critical role in the subsequent tasks performed by participants in the present study. Individuals with low-level CSEs at the onset of a search task demonstrate higher-level, inflated CSEs once they gain access to search tools.  Therefore, knowing the source of CSE measured before a search activity becomes critical at this point. Our findings point out the need to measure individuals’ CSEs at the onset of a search task and to consider potential remedies that can reduce inflated CSE and mitigate its unintended consequences. Several targeted interventions could address these challenges by promoting more realistic self-assessments and balanced reliance on external memory systems. First, search tool usage training could equip users with critical evaluation strategies, helping them distinguish between their own knowledge and tool-facilitated successes. Second, reflective exercises, such as asking users to identify what they knew versus what they learned through searches, could enhance self-awareness of cognitive reliance. Additionally, feedback mechanisms embedded within search platforms could prompt users to reflect on their dependence on external tools. For instance, prompts could remind users that certain successes were tool-assisted and encourage them to consider their ability to replicate these successes independently. Finally, educational and professional tasks could be designed to include both independent and tool-assisted phases, fostering a balance between autonomous thinking and effective tool use. These interventions aim to mitigate inflated CSE by promoting realistic self-assessments, improving critical thinking, and encouraging a more balanced approach to leveraging external tools. 

Beyond the potential remedies discussed, these findings underscore the significant role of search systems in shaping users’ cognitive self-esteem and their reliance on external memory systems. As the internet increasingly functions as a memory partner, search engines become critical interfaces that mediate this relationship. The interplay between search self-efficacy, cognitive offloading, and evaluations of cognitive abilities suggests several implications for designing search systems to better support users' cognitive processes while mitigating potential risks. For example, features that nudge users to recall related information from their memory before presenting search results could address the blurring effect between internal and external knowledge. Furthermore, supporting users in building confidence and competence in their search abilities can be prioritized. Real-time feedback on query formulation and suggestions for refining search strategies can enhance users’ perceptions of their search skills and foster long-term development of effective search strategies. This is particularly important for users with lower initial CSE, as their cognitive evaluations are significantly shaped by search self-efficacy. However, excessive reliance on external systems, as evidenced in individuals with lower CSE who blur the boundaries between their memory and search systems, highlights the importance of designing systems that also encourage internal cognitive processing. Transparency features that differentiate information sourced by the user versus the system can help mitigate overestimation of one’s own knowledge. 

Findings of the present study may have an impact on how search tools are used in educational settings. To date, several educational interventions such as manipulatives \cite{martin2005physically, pouw2014embedded} or calculators \cite{ellington2003meta,hembree1986effects} have assisted students to reduce the extra mental effort required to learn new information (e.g., performing computational tasks). Such interventions have been effective in cognitive offloading to ensure that the cognitive load does not exceed students’ processing capacity, so they will not struggle to complete an activity successfully \cite{paas2003cognitive}. Currently, search engines play a critical role in higher education, and their use is prevalent among students \cite{jadhav2011significant}. Students tend to offload the responsibility of storing and retrieving information to search tools, and such an offloading strategy leads to concerns about the quality of students’ learning. Our findings suggest search engines can blur boundaries between what students really know and what they believe to know when they use search engines as cognitive offloading tools. Given that using search tools to access information has become a norm in today’s society, some strategies should be developed to address raising concerns without prohibiting students from using search tools in the classroom. We can benefit from the slow search concept that advocates reducing speed in favor of increasing quality.  The slow search was introduced as a way to give search systems additional time to provide a higher quality search experience in exchange of reducing speed in retrieving search results \cite{teevan2013slow}. Research suggests that even minor delays in the response time of web search engines can result in a significant reduction in user engagement\cite{brutlag2009speed}. In addition, delivering results slowly can lead to a dramatic drop in the perceived quality of results \cite{teevan2013slow}. Thus, introducing structured delays or slow search mechanisms could prompt users to engage more with their internal memory resources, reducing reliance on search tools for immediate answers.

\subsection{Limitations of the study  }

While this study provides valuable insights into the relationship between search tools and cognitive self-esteem (CSE), it is not without its limitations. First, the study was conducted exclusively with undergraduate students from an information technology program. This sampling approach limits the generalizability of the findings to broader populations. University students, especially those studying in technology-related fields, are likely to have higher familiarity with search tools and may experience unique pressures around information retrieval and self-evaluation compared to individuals in non-academic or non-technical settings. Future research should consider including participants from diverse demographics, such as older adults, individuals in non-academic environments, or those with different professional expertise. For instance, individuals engaging in physical, real-world problem-solving may exhibit different patterns in how they frame CSE when interacting with search systems.

Second, the study’s scope was restricted to general knowledge questions with low recall probabilities to ensure the reliability of results. However, different types of search tasks or domains, such as problem-solving, creative tasks, or domain-specific research, might elicit varying impacts on CSE. Examining whether the observed patterns hold for more complex or less structured search tasks could provide a more nuanced understanding of the interaction between search tools and cognitive evaluations.

Third, the difficulty level of the tasks was standardized, but this does not account for how varying levels of task complexity could influence participants’ cognitive evaluations. For example, more challenging questions might amplify the reliance on external search tools, while simpler questions might favor internal memory retrieval. Future studies should manipulate task difficulty systematically to explore how it moderates the relationship between search tool access and CSE.

Finally, the study was conducted over a relatively short timeframe with only four repeated measures. As suggested by Risko and Gilbert’s model of cognitive offloading, longer timeframes and a greater number of repeated measures could provide deeper insights into longitudinal changes in CSE and offloading behaviors. Expanding the study to track participants’ cognitive evaluations over extended periods or across multiple contexts could further illuminate the durability and variability of these effects.

\subsection{Future Research Directions}
Future studies should investigate both the implications of inflated CSE due to search engine access and the long-term dynamics of fluctuating CSE levels. For instance, longitudinal research could explore whether inflated CSE persists with repeated reliance on search tools, and how this phenomenon influences learning outcomes, critical thinking, and decision-making. Such studies could also address whether these effects vary across different populations or depend on the nature of the tasks performed. A longer-term longitudinal design could explore whether the observed effects on cognitive self-esteem (CSE) persist, diminish, or evolve over extended periods, offering deeper insights into the interplay between situational interventions and trait-level stability. This would provide clarity on the extent to which the observed CSE fluctuations are transient or reflect more enduring cognitive shifts.

Interventions aimed at mitigating overconfidence, such as training on effective information evaluation and retention strategies, could also be explored to counterbalance the potential downsides of inflated CSE. Additionally, examining the role of individual differences, such as baseline CSE levels and search self-efficacy, could provide a more nuanced understanding of how these factors interact with the accessibility of external memory systems. Future research could control for device familiarity by using standardized or mixed-device setups, which could isolate its role in shaping CSE. For instance, participants could be required to alternate between familiar and unfamiliar devices, enabling researchers to disentangle the effects of device familiarity from task-specific interventions. Given that our study observed differences in device usage (personal devices vs. classroom computers), this methodological refinement would provide deeper insights into how cognitive self-esteem is shaped by familiarity with technological tools.

Expanding the participant pool to include diverse demographics, such as older adults or those with limited technological literacy, would also enhance the generalizability of findings. University students, particularly those in technology-oriented disciplines, may have higher baseline familiarity with search tools, which could influence the effects on cognitive self-esteem. By including more demographically diverse populations, researchers could determine the extent to which the observed patterns are context-dependent or generalizable to broader populations.

Finally, researchers should examine how emerging technologies like generative AI and voice-based search systems might affect cognitive self-esteem. Unlike traditional search tools, AI-driven systems may offer users more interpretative support, suggesting that users might attribute greater cognitive ability to themselves when using such tools. Voice-based assistants like Siri or Alexa might further influence user perceptions, as they reduce the need for manual query construction, potentially altering the user's perception of their own cognitive involvement. Interventions such as search training or structured feedback could be tested to amplify the positive impacts of self-efficacy on CSE. For example, prompting users to reflect on their reliance on search tools and encouraging independent recall of key information may enhance users’ cognitive self-awareness and reduce overreliance on search engines as cognitive partners.

\section{Conclusion}
The present study contributed to the literature by providing evidence that accessing search tools does not impact individuals’ CSEs in the same way, with some who showed a noticeable change in their CSE after gaining access to search tools, and those who exhibited very little change in their CSE even after they gained access. The search self-efficacy demonstrated here represents one factor that impacts individuals’ CSEs differently. Individuals who reported relatively higher CSEs did not demonstrate a noticeable change throughout the study, and search self-efficacy did not impact their CSEs significantly. On the other hand, individuals who reported relatively lower CSEs exhibited an important change in their CSEs, and they were impacted significantly by search self-efficacy. These results contribute to our understanding of how cognitive evaluations are influenced in a human-technology transactive system. 

\bibliographystyle{unsrt}  
\bibliography{references} 
\end{document}